\RequirePackage{fix-cm}
\documentclass[smallextended]{svjour3}       % onecolumn (second format)                                                
\smartqed  % flush right qed marks, e.g. at end of proof                                                                
\usepackage{graphicx}
\usepackage{bm}
 \usepackage{amsfonts}
\usepackage{amsmath}
\usepackage{amssymb}                                                                     

\begin{document}

\title{Efimov effect evaporation after confinement\thanks{This work has been partially supported by
the Spanish Ministry of Science, Innovation and University
MCIU/AEI/FEDER,UE (Spain) under Contract No. PGC2018-093636-B-I00.}                                      
}


\author{E. Garrido         \and A.S. Jensen                                                                                           
}


\institute{E. Garrido \at
               Instituto de Estructura de la Materia, CSIC, Serrano 123, E-28006 Madrid, Spain\\
              \email{e.garrido@csic.es}           %  \\                                                               
           \and
           A.S: Jensen\at
              Department of Physics and Astronomy, Aarhus University, DK-8000 Aarhus C, Denmark
}

\date{Received: date / Accepted: date}

\maketitle

%%%%%%%%%%%%%%%%%%%%%%%%%%%%%%%%%%%%%%%%%%%%%%%%%%%%%%%%%%%%%%%%%%%%

\begin{abstract}
The continuous confinement of quantum systems can be described by means of the $d$-method, where the dimension $d$ is 
taken as a continuous parameter. In this work we describe in detail how this method can be used to obtain the root mean
square radii for a squeezed three-body system. These observables are used  to investigate the disappearance of the Efimov states around the
two-body threshold during a progressive confinement of the system from three to two dimensions. We illustrate how the disappearance takes 
place through the loss of one of the particles, whereas the other two remain bound.
\end{abstract}

\maketitle

\paragraph{\it Introduction.}
The properties of few-body systems are to a large extent determined by the dimension of the space where they are allowed to
move. The simplest example is perhaps the case of a two-body system,  always bound in two dimensions (2D) by any infinitesimally small
attractive potential, but not necessarily bound in three dimensions (3D) due to the different behavior of the centrifugal barrier \cite{sim76,lan77}.
This fact has two immediate consequences. The first one is that the Efimov effect, present in three-body systems where at least 
two of the three pair-interactions have nearly zero energy \cite{efi70}, does not exist in 2D \cite{nie97,vol13,nai17,bru79,lim80}.
The second one is that, when confining an unbound two-body system from 3D to 2D, there must necessarily be a point in the
confinement process where the two-body system is precisely at the zero-energy threshold. 

From this last result one can also conclude that, given an unbound three-body system containing at least two identical
particles, when confining from 3D to 2D there will always be a point in the confinement process where the Efimov conditions are fulfilled. This fact
has  recently been investigated in Ref.~\cite{gar21}, where the confinement of the unbound three-body state was implemented by using the dimension $d$
as a parameter that moves continuously from $d=3$ to $d=2$, in such a way that for some particular value, $d=d_E$, the Efimov
effect shows up. One of the main results in \cite{gar21} is that the Efimov states appear and disappear around the two-body threshold extremely fast
within  a very small dimension interval around $d=d_E$.

The equivalence between the $d$-method, using $d$ as a continuous parameter, and the more natural approach of confining the system by
means of an external squeezing potential included explicitly in a 3D calculation,  has been investigated in Refs.~ \cite{gar19,gar19b,gar20}.
The main conclusion is that it is possible to establish a connection between a non-integer value of $d$ and a specific strength
of the squeezing potential. In this way, the $d$-method, much simpler to implement numerically,  appears as an efficient tool 
to describe confined systems.

The price to pay when using the $d$-formalism is that the calculation of a given observable, unavoidably measured in an integer-dimensional space, 
requires a  translation of the $d$-dimensional wave function into the ordinary 3D space. This is done by considering 
the $d$-wave function as a wave function in the ordinary 3D space, but deformed along the squeezing direction \cite{gar19,gar19b,gar20}.

The purpose of this letter is twofold. First, to clarify how the three-body wave functions obtained within the $d$-method can in practice
be used to compute observables, in particular root mean square radii, which are very relevant in order to understand the spatial distribution
of the three constituents. The squeezing of the system along the $z$-axis (3D to 2D squeezing) will be considered. And second, to employ 
these observables to visualize how, during the squeezing process from 3D to 2D, the 
disappearance of the Efimov states close to the two-body threshold does actually take place. 

The results will be illustrated with one of the systems detailed in Ref.~\cite{gar21}, namely the case of two heavy and one light particles
 with mass ratio  $m_H/m_L=133/6$, and with heavy-light potential such that $d_E=2.75$.

\paragraph{\it Three-body mean-square radii with the $d$-method.}
A three-body system in a $d$-dimensional space has, after removal of the center of mass motion, $2d$ degrees of freedom, which can
be described by means of the well-known Jacobi coordinates, $\bm{x}$ and $\bm{y}$. 
As usual, from these coordinates one can
construct the hyperradius, $\rho=(x^2+y^2)^{1/2}$, and the hyperangle, $\alpha =\arctan(x/y)$, which, 
together with the $2(d-1)$ angles defining the directions of $\bm{x}$ and $\bm{y}$, constitute the $2d$ hyperspherical coordinates
necessary to describe the system. 

Using these coordinates, the $d$-dimensional wave function can be obtained by means of
the hyperspherical adiabatic expansion method, described in detail in Ref.~\cite{nie01}, in such a way that the wave function is
written as:
\begin{equation}
\Psi_d(\rho,\Omega_d)=\frac{1}{\rho^{\frac{2d-1}{2}}} \sum_n f_n^{(d)}(\rho) \Phi_n^{(d)}(\rho, \Omega_d),
\label{eq2}
\end{equation}
where $\Omega_d$ collects the $2d-1$ hyperangles. The angular functions $\Phi_n^{(d)}(\rho, \Omega_d)$ are the eigenfunctions
of the angular part of the Faddeev (or Schr\"{o}dinger) equations, whereas the functions $f_n^{(d)}(\rho)$ are obtained after
solving a coupled set of radial, Schr\"{o}dinger-like, differential equations where the eigenvalues of the angular part enter as
effective potentials.

As already mentioned, the use Eq.~(\ref{eq2}) in order to compute whatever observable, requires a  translation  of the
$\Psi_d$ wave function into the 3D space in which the observable is measured. This connection is done
 by considering the $d$-wave function (\ref{eq2}) as a wave function in the ordinary 3D space, but deformed along the squeezing direction (that we choose along the $z$-axis). In other words,
the Jacobi coordinates, $\bm{x}$ and $\bm{y}$ in the $d$-space, will be taken as ordinary 3D vectors, written as $\tilde{\bm{x}}$ and $\tilde{\bm{y}}$. Each of them
has therefore the usual Cartesian components, where $\tilde{x}_z$ and $\tilde{y}_z$ shall denote the corresponding components along the $z$-axis, and
$\tilde{x}_\perp$ and $\tilde{y}_\perp$ the projection of $\tilde{\bm{x}}$ and $\tilde{\bm{y}}$ on the $xy$-plane.

The key point is that in order to incorporate the deformation produced by the external field, the $\tilde{x}_z$ and $\tilde{y}_z$ components  along 
the squeezing direction do not represent the actual value of the coordinate, but instead, the actual value deformed by means of a scale
parameter $s$, i.e., $\tilde{x}_z=x_z/s$ and  $\tilde{y}_z=y_z/s$. In the same way, the perpendicular components along the non-squeezed
directions do provide the true value along these directions, i.e., $\tilde{x}_\perp=x_\perp$ and  $\tilde{y}_\perp=y_\perp$. 
Therefore, for squeezing along the $z$-axis we can write:
\begin{equation}
\tilde{x}^2=\tilde{x}_\perp^2+\tilde{x}_z^2=x_\perp^2+\frac{x_z^2}{s^2},
\label{tilx}
\end{equation}
\begin{equation}
\tilde{y}^2=\tilde{y}_\perp^2+\tilde{y}_z^2=y_\perp^2+\frac{y_z^2}{s^2},
\label{tily}
\end{equation}
where the scale parameter $s$ can take values within the range $0 \leq s \leq 1$. For $s=0$ only the values $x_z=0$ and $y_z=0$ are allowed (otherwise the 
value of the coordinates diverges), and the system is fully squeezed into the 2D plane. For $s=1$ we have $\tilde{x}^2=x_\perp+x_z^2$ and   $\tilde{y}^2=y_\perp+y_z^2$, and
$\tilde{\bm{x}}$ and $\tilde{\bm{y}}$ become the usual coordinates in a non-squeezed space. Equations (\ref{tilx}) and (\ref{tily}) can be easily
generalized to other squeezing scenarios, like from 3 to 1 dimensions, or from 2 to 1 dimensions \cite{gar19b}.

Using the 3D $\tilde{\bm{x}}$ and $\tilde{\bm{y}}$ Jacobi coordinates we can construct the usual 3D hyperspherical coordinates, which together with the angles 
$\Omega_{\tilde{x}} \equiv \{\theta_{\tilde{x}}, \varphi_{\tilde{x}} \}$ and $\Omega_{\tilde{y}} \equiv \{ \theta_{\tilde{y}}, \varphi_{\tilde{y}} \}$ giving the directions of $\tilde{\bm{x}}$ and $\tilde{\bm{y}}$, contain
the hyperradius $\tilde{\rho}$ and the hyperangle $\tilde{\alpha}$:
\begin{equation}
\tilde{\rho}^2=\tilde{x}^2+\tilde{y}^2, \hspace*{1cm} \tilde{\alpha}=\arctan \left( \frac{\tilde{x}}{\tilde{y}} \right),
\label{eq5}
\end{equation}
or, equivalently, 
\begin{equation}
\tilde{x}=\tilde{\rho}\sin\tilde{\alpha}, \hspace*{1cm} \tilde{y}=\tilde{\rho}\cos\tilde{\alpha}.
\label{hyp2}
\end{equation}

Also, from the definitions above it is simple to see that the ordinary 3D volume element can be written as:
\begin{equation}
dV=x_\perp dx_\perp dx_z d\varphi_x  y_\perp dy_\perp dy_z d\varphi_y=
s^2 \tilde{x}_\perp d\tilde{x}_\perp d\tilde{x}_z d\varphi_{\tilde{x}}  \tilde{y}_\perp d\tilde{y}_\perp d\tilde{y}_z d\varphi_ {\tilde{y}}=s^2 d\tilde{V}.
\label{eq7}
\end{equation}

Once the $d$-wave function $\Psi_d$ in Eq.(\ref{eq2}) is interpreted as a deformed wave function in the ordinary 3D space, it is necessary to normalize
correctly the wave function in three dimensions. Thanks to the relation above we can obtain the normalization constant as: 
\begin{equation}
C(s)^2=\int dV \left| \Psi_d(x_\perp, x_z, y_\perp, y_z,s) \right|^2=s^2 \int d\tilde{V} \left| \Psi_d(\tilde{\bm{x}}, \tilde{\bm{y}})\right|^2,
\label{cs}
\end{equation} 
where the last integral does not depend on $s$.

The wave function $\tilde{\Psi}_d=\Psi_d/C(s)$ is correctly normalized to 1 in the ordinary 3D space, and it can therefore be employed to compute
the expectation value of whatever observable of interest. In particular, we shall focus here on $\langle x^2 \rangle$, which is given by:
\begin{equation}
\langle x^2\rangle= \int dV x^2 \left| \tilde{\Psi}_d(x_\perp, x_z, y_\perp, y_z, s)\right|^2=\int dV (x_\perp^2+x_z^2) \left| \tilde{\Psi}_d(x_\perp, x_z, y_\perp, y_z, s)\right|^2,
\end{equation}
which by use of Eq.(\ref{eq7}), and keeping in mind that $\tilde{x}_\perp=x_\perp$ and $\tilde{x}_z=x_z/s$, can be written as:
\begin{equation}
\langle x^2\rangle= 
s^2 \int d\tilde{V} \tilde{x}_\perp^2 \left| \tilde{\Psi}_d(\tilde{\bm{x}},\tilde{\bm{y}})\right|^2+s^4 \int d\tilde{V} \tilde{x}_z^2 \left| \tilde{\Psi}_d(\tilde{\bm{x}},\tilde{\bm{y}})\right|^2.
\end{equation}

Since $\tilde{\Psi_d}=\Psi_d/C(s)$ and $C(s)^2$ is given by Eq.(\ref{cs}),  the equation above becomes:
\begin{equation}
\langle x^2\rangle=\frac{
 \int d\tilde{V} \tilde{x}_\perp^2 \left| \Psi_d(\tilde{\bm{x}},\tilde{\bm{y}})\right|^2+s^2 \int d\tilde{V} \tilde{x}_z^2 \left| \Psi_d(\tilde{\bm{x}},\tilde{\bm{y}})\right|^2
 }{\int d\tilde{V} \left| \Psi_d(\tilde{\bm{x}}, \tilde{\bm{y}})\right|^2}.
 \label{eq11}
\end{equation}

The integrals in Eq.(\ref{eq11}) do not depend on $s$. This actually implies that, in terms of the $\tilde{\bm{x}}$ and $\tilde{\bm{y}}$ coordinates, there is no preferred direction 
in the 3D space, and we therefore have that:
\begin{equation}
\int d\tilde{V} \tilde{x}_i^2 \left| \Psi_d(\tilde{\bm{x}},\tilde{\bm{y}})\right|^2= 
\frac{1}{3} \int d\tilde{V} \tilde{x}^2 \left| \Psi_d(\tilde{\bm{x}},\tilde{\bm{y}})\right|^2,
\label{eq12}
\end{equation}
where $i$ is any of the Cartesian components of the $\tilde{\bm{x}}$ vector in the 3D space.

Using Eq(\ref{eq12}) we can then finally write:
\begin{equation}
\langle x^2\rangle= \frac{2+s^2}{3}  \frac{\int d\tilde{V} \tilde{x}^2 \left| \Psi_d(\tilde{\bm{x}},\tilde{\bm{y}})\right|^2}
{\int d\tilde{V} \left| \Psi_d(\tilde{\bm{x}}, \tilde{\bm{y}})\right|^2},
\end{equation}
and similarly:
\begin{equation}
\langle y^2\rangle= \frac{2+s^2}{3}  \frac{\int d\tilde{V} \tilde{y}^2 \left| \Psi_d(\tilde{\bm{x}},\tilde{\bm{y}})\right|^2}
{\int d\tilde{V} \left| \Psi_d(\tilde{\bm{x}}, \tilde{\bm{y}})\right|^2}.
\end{equation}

%In hyperspherical coordinates, we have that:
%\begin{equation}
%d\tilde{V}=\tilde{\rho}^5 d\tilde{\rho} \sin^2\tilde\alpha \cos^2\tilde{\alpha} d\tilde\alpha d\Omega_{\tilde{x}} d\Omega_{\tilde{y}}
%\end{equation}
%where $\Omega_{\tilde{x}}$ and $\Omega_{\tilde{y}}$ contain the usual polar and azimuthal angles describing the direction of $\tilde{\bm{x}}$ and $\tilde{\bm{y}}$, respectively.

When only $s$-waves are included in the calculation, the wave function $\Psi_d$ does not depend on  $\Omega_{\tilde{x}}$ and $\Omega_{\tilde{y}}$, and making use of 
Eqs.(\ref{eq5}) and (\ref{hyp2}) we can write:
\begin{equation}
\langle x^2\rangle= \frac{2+s^2}{3}  \frac{\int \tilde{\rho}^7 d\tilde{\rho} \sin^4\tilde\alpha \cos^2\tilde{\alpha} d\tilde\alpha  \left| \Psi_d(\tilde{\rho},\tilde{\alpha})\right|^2}
{\int \tilde{\rho}^5 d\tilde{\rho} \sin^2\tilde\alpha \cos^2\tilde{\alpha} d\tilde\alpha \left| \Psi_d(\tilde{\rho}, \tilde{\alpha})\right|^2}
\end{equation}
and
\begin{equation}
\langle y^2\rangle= \frac{2+s^2}{3}  \frac{\int \tilde{\rho}^7 d\tilde{\rho} \sin^2\tilde\alpha \cos^4\tilde{\alpha} d\tilde\alpha  \left| \Psi_d(\tilde{\rho},\tilde{\alpha})\right|^2}
{\int \tilde{\rho}^5 d\tilde{\rho} \sin^2\tilde\alpha \cos^2\tilde{\alpha} d\tilde\alpha \left| \Psi_d(\tilde{\rho}, \tilde{\alpha})\right|^2}.
\end{equation}

Note that the ratio
\begin{equation}
\frac{ \langle y^2 \rangle }{\langle x^2 \rangle}=\frac{\int d\tilde{V} \tilde{y}^2 \left| \Psi_d(\tilde{\bm{x}},\tilde{\bm{y}})\right|^2}
{\int d\tilde{V} \tilde{x}^2 \left| \Psi_d(\tilde{\bm{x}},\tilde{\bm{y}})\right|^2}=
\frac{\int \tilde{\rho}^7 d\tilde{\rho} \sin^2\tilde\alpha \cos^4\tilde{\alpha} d\tilde\alpha \left| \Psi_d(\tilde{\rho},\tilde{\alpha})\right|^2}
{\int \tilde{\rho}^7 d\tilde{\rho} \sin^4\tilde\alpha \cos^2\tilde{\alpha} d\tilde\alpha  \left| \Psi_d(\tilde{\rho},\tilde{\alpha})\right|^2}
\label{eq17}
\end{equation}
does not depend on the scale parameter.

The Jacobi coordinates do not provide the actual relative coordinates between  the particles, $\bm{r}_x$ and $\bm{r}_y$, but they are related through proportionality factors
depending on the mass of the particles \cite{nie01}. Taking this into account, one can easily get:
\begin{equation}
\frac{ \langle r_y^2 \rangle^{1/2} }{\langle r_x^2 \rangle^{1/2}}=\sqrt{ \frac{m_k(m_i+m_j) }{m_i+m_j+m_k }  \frac{m_i+m_j }{m_i m_j } } 
\frac{ \langle y^2 \rangle^{1/2}}{\langle x^2 \rangle^{1/2}},
\label{dist}
\end{equation}
where $r_x$ is the distance between the particles $i$ and $j$, connected by the $x$-coordinate, and $r_y$ is the distance between particle $k$ and the center of mass of
the $ij$-system, which are connected by the $y$-coordinate.

\paragraph{\it Results.}

We shall consider the case of two identical spinless heavy particles with mass $m_H$ and one spinless light particle with mass $m_L$. Only relative $s$-waves will be included in the
calculation. The interaction between the two heavy particles is put equal to zero.
The interaction between the light and the heavy particles is such that the heavy-light two-body system and the three-body system are both unbound in three dimensions. 

As mentioned, it is known that in two dimensions any infinitesimal attraction is able to support at least one bound two-body state \cite{sim76,lan77}. Therefore, along the squeezing process,
from three to two dimensions,  the heavy-light system unavoidably becomes bound for some particular strength of the external squeezing potential, or, equivalently, within the $d$-formalism, for some specific value
of the non-integer dimension, $d=d_E$. For this dimension, $d_E$,  the two-body binding energy is equal to zero, and, since only relative $s$-waves are considered, the Efimov conditions
are fulfilled \cite{gar21}. 

The energy separation between the Efimov states depends crucially on the $m_H/m_L$ mass ratio, in such a way that the larger the ratio, the closer the Efimov states. In here we
shall use the heavy-light Gaussian potential specified in Ref.~\cite{gar21} (the range of the potential is taken as length unit), which determines the appearance of the Efimov states for $d=d_E=2.75$. 
We shall consider the mass ratio $m_H/m_L=133/6$, which, as shown in \cite{gar21}, corresponds to an energy scale factor between the Efimov states of $46.5$.

In Ref.~\cite{gar21} it is also shown (Fig.~5 in \cite{gar21}) that, when squeezing to dimensions smaller than $d_E$, the three lowest Efimov states survive as bound three-body
states down to $d=2$ \cite{bel13}. The other infinitely many Efimov states disappear very fast around the two-body threshold. This fact  was interpreted as a process where, when decreasing $d$, one 
of the heavy particles is becoming less and less bound with respect to the bound heavy-light two-body system, being eventually lost \cite{gar21,gar20}.

\begin{figure}[t!]
\begin{center}
\includegraphics[width=\textwidth,angle=0]{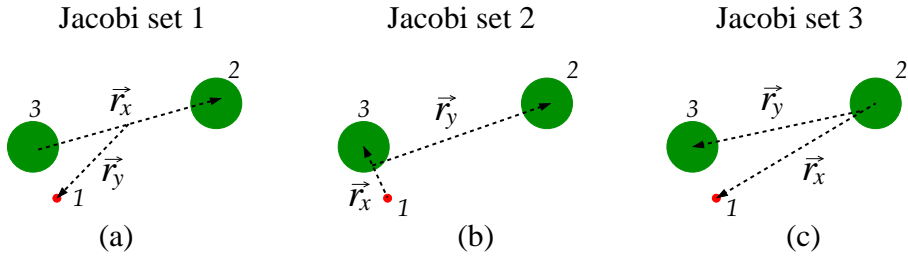}
\end{center}
\caption{Definition of the three possible Jacobi sets. }
\label{fig1}
\end{figure}

Making use of Eqs.(\ref{eq17}) and (\ref{dist}) one can easily visualize how the disappearance process takes place. To do so, we define the three possible Jacobi sets as shown in Fig.~\ref{fig1}.
If the interpretation made in \cite{gar21,gar20} is correct, when close to the state disappearance, one of the heavy particles, particle 2 in Fig.~\ref{fig1}, should be pretty far from the bound heavy-light
system (particles 1 and 3). This means that, when close to the disappearance of the state, if the first Jacobi set is used (Fig.~\ref{fig1}a) we should get $r_y/r_x \approx 0.5$ (since 
the two heavy particles are taken identical, the center of mass of the heavy-heavy system is at half the distance between them). In the same way, if the second Jacobi set is employed, Fig.~\ref{fig1}b, 
we should get $r_y/r_x \rightarrow \infty$. And finally, in the third Jacobi set, Fig.~\ref{fig1}c, we should obtain $r_y/r_x \approx 1$, since the center of mass of the heavy-light system is
pretty close to the center of mass of the heavy particle. 

Note however that, since the two heavy particles are identical, the second and third Jacobi sets are indistinguishable. Therefore, the use of any of them will simultaneously contain the features 
corresponding to the two Jacobi sets. The consequence is that, when using the second, or third, Jacobi set to describe the three-body system, only the divergence of the $r_y/r_x$ ratio  will be visible. 

\begin{figure}[t!]
\begin{center}
\includegraphics[width=8cm,angle=0]{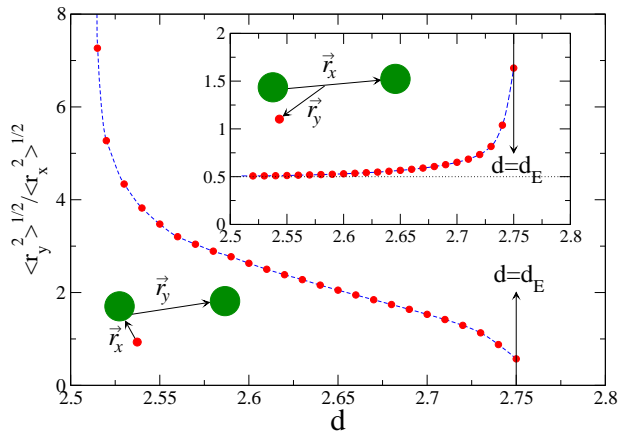}
\end{center}
\caption{The outer panel shows, as a function of $d$, the ratio $\langle r_y^2\rangle^{1/2} /\langle r_ x^2 \rangle^{1/2}$ when computed in the second Jacobi set, as depicted in Fig.~\ref{fig1}b. The inset
shows the same ratio when computed in the first Jacobi set,  Fig.~\ref{fig1}a. The arrows indicate value of the dimension $d_E$ for which the Efimov conditions are fulfilled. }
\label{fig2}
\end{figure}

To check the results discussed in the previous paragraphs, we consider the third excited Efimov state for the three-body system under consideration (the $m_H/m_L=133/6$ case in Ref.\cite{gar21}). This is 
the last of the Efimov states disappearing during the squeezing process for $d<d_E$. For this state, we show in Fig.~\ref{fig2}  the ratio $\langle r_y^2\rangle^{1/2}/\langle r_x^2\rangle^{1/2}$, Eq.(\ref{dist}), as a
 function of the dimension $d$. The curves in the inner and outer panels,  show, respectively,  the result when the first and second Jacobi sets are used to describe the system. As we can
see, for $d=d_E=2.75$, where the Efimov effect takes place, the radius ratios take values of $\sim 1.64$ and $\sim 0.57$ in the first and second Jacobi sets, respectively. These values are consistent with
an isosceles triangular spatial geometry, where the light particle is located in the vertex connecting the two equal sides of the triangle. The distance between each of the  heavy particles and 
the light one is about 1.75 times the distance between the two heavy particles.

When reducing the dimension, the geometry changes, approaching more and more the one depicted in Fig.~\ref{fig1}, with the light particle close to one  of the heavy ones, and the second heavy particle
moving far apart. In fact, as we can see in Fig,~\ref{fig2}, when close to the dimension at which the Efimov states disappears ($d\approx 2.51$), the curve corresponding to the first (inner panel) and second 
(outer panel) Jacobi sets approaches the value of 0.5 and diverges to $+\infty$, respectively. As mentioned when discussing Fig.~\ref{fig1}, this is the fingerprint of the disappearance process anticipated in Refs.~\cite{gar21,gar20}, namely,
one of the heavy particles is moving far apart, being progressively less and less bound with respect to the bound heavy-light system, and lying eventually into the continuum. 

\begin{figure}[t!]
\begin{center}
\includegraphics[width=\textwidth]{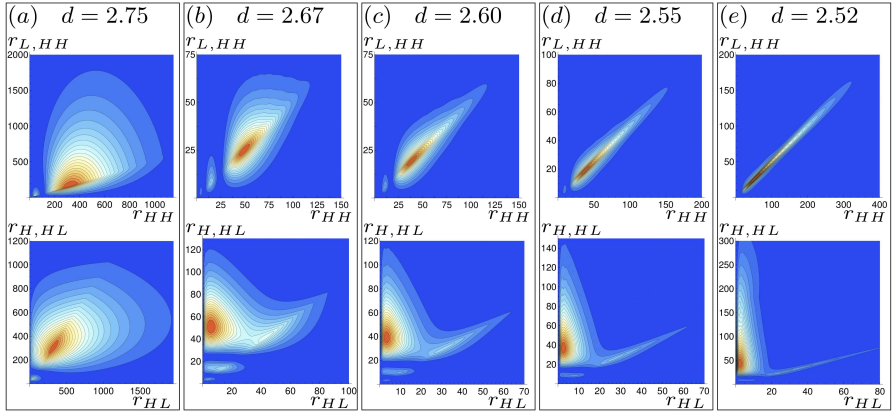}
\end{center}
\caption{Contour plots of the density function in Eq.(\ref{densfun}) for dimension values $d=2.75$, 2.67, 2.60, 2.55, and 2.52, panels (a), (b), (c), (d), and (e), respectively. On each panel the upper and
lower plots correspond to the calculations in the first and second Jacobi sets, respectively. The lengths are given in units of the range of the Gaussian interaction between the two heavy particles.}
\label{fig3}
\end{figure}

This behaviour can also be seen in Fig.~\ref{fig3}, where we show, for different values of the dimension $d$, the contour plots of the density function 
\begin{equation}
F(r_x,r_y) = \int r_x^2 r_y^2 d\Omega_{x} d\Omega_{y} \left| \tilde{\Psi}_d(\bm{x},\bm{y},s) \right| ^2,
\label{densfun}
\end{equation}
where, for simplicity, since it does not affect our illustration purpose, we have taken $s=1$ for all the values of $d$. 

In the figure, the panels (a), (b), (c), (d), and (e) correspond to the dimension $d=d_E=2.75$, $d=2.67$, $d=2.60$, $d=2.55$, and $d=2.52$,
respectively. In all the panels the upper figure shows the result obtained in the first Jacobi set,
where $r_x=r_{HH}$ and $r_y=r_{L,HH}$, respectively, whereas the lower figure is
the result in the second Jacobi set, where $r_x=r_{HL}$ and $r_y=r_{H,HL}$. The distances are
given in units of the range of the Gaussian interaction between the two heavy particles.

In the upper part of Fig.~\ref{fig3}a, which corresponds to $d=d_E$, we can see that, although the density function is peaked at value matching the ratio $r_{L,HH}/r_{HH}\approx 0.5$, the
wave function shows however a long tail, mainly in the $r_{L,HH}$ direction, which increases pretty much the expectation value of $\langle r_{L,HH}^2 \rangle$. This leads to the ratio 
$\langle r_{L,HH}^2 \rangle^{1/2}/\langle r_{HH}^2 \rangle^{1/2}\approx 1.64$ shown in the inner part of Fig.~\ref{fig2} for $d=d_E$. When squeezing the system (decreasing $d$), we can 
observe, by following the upper figures in the five panels from left to right, that the tail of the wave function disappears progressively, in such a way that for $d=2.52$ (upper part of Fig.~\ref{fig3}e),
close to the disappearance of the state, the wave function reduces to a very narrow stripe that follows very closely the axis $r_{L,HH}=r_{HH}/2$, very consistent with the geometry shown
in Fig.,{\ref{fig1}a, where one of the heavy particles is very far from the bound heavy-light system.

A similar conclusion is reached when analyzing the lower plots in Fig.~\ref{fig3}, which correspond to the results in the second Jacobi set. For $d=d_E=2.75$ (lower plot in Fig.~\ref{fig3}a)
the wave function is peaked at a value of $r_{H,HL}$ a bit smaller than $r_{HL}$, consistent with the isosceles structure where the two heavy particles are closer to each other than
the light and the heavy particles (keep in mind that $r_{H,HL}\approx r_{HH}$). Furthermore, the wave function is clearly more extended along the $r_{HL}$-axis than along 
the vertical axis, which leads to the ratio $\langle r_{H,HL}^2 \rangle^{1/2}/\langle r_{HL˘}^2 \rangle^{1/2}\approx 0.57$ shown in the outer part of Fig.~\ref{fig2} for $d=d_E$.
When squeezing the system, as we can see when moving to the right in the lower plots in Fig.~\ref{fig3}, the peak is moving towards small values of $r_{HL}$, and, at the same time, the central 
part of the density distribution is progressively disappearing. The result is that, when close to the disappearance of the state, $d=2.55$ and $2.52$, it can be clearly seen that the density
function presents two very distinct components, one of them with $r_{HL}\ll r_{H,HL}$, consistent with the structure shown in Fig.~\ref{fig1}b, and a second component with $r_{HL}\approx r_{H,HL}$,
consistent with the structure in Fig.~\ref{fig1}c. The simultaneous presence of these two components is a consequence of the fact that the second and third Jacobi sets are indistinguishable.

\paragraph{\it Conclusions.}

In previous works the $d$-method was shown to be an efficient tool in order to describe squeezed two- and three-body systems, essentially due to the fact that the external squeezing 
potential does not enter explicitly \cite{gar21,gar19,gar19b,gar20}. This fact strongly simplifies the calculation of the wave functions, but at the expense of complicating the calculation of 
expectation values, since the wave function in $d$ dimensions has to be translated into the ordinary three-dimensional space.

In this work we have focused on three-body systems squeezed from three to two dimensions,  showing how the corresponding root mean square radii can be extracted after interpretation of the 
$d$-wave function as a 3D wave function deformed along the squeezing direction. This procedure can be similarly applied to other observables, as well as other squeezing scenarios, like from 
three to one, or from two to one, dimensions. The computed radii have been then used to  investigate how the Efimov states disappear around the two-body threshold during the
squeezing process.

We have considered a three-body system made of two identical heavy particles and a light particle. Neither the three-body system, nor any of the two-body subsystems, is bound in 3D. 
As shown in \cite{gar21}, during the squeezing process the system must, for some particular squeezing ($d=d_E$),
 fulfill the Efimov conditions, but, except a few of them that survive
down to $d=2$, the rest of the Efimov states disappear very fast around the two-body threshold. In this work we have visualized how this disappearance process takes place. We have shown
that this happens through the loss of one of the heavy particles, which becomes progressively less and less bound with respect to the bound heavy-light system, being then eventually
released into the continuum.

\end{document}